\documentclass[conference,a4paper]{IEEEtran}
\usepackage{geometry}
\usepackage{xcolor}
\usepackage{balance}
\usepackage{multirow}

\ifCLASSINFOpdf
  \usepackage[pdftex]{graphicx}
\else
  \usepackage[dvips]{graphicx}
\fi

\usepackage{xcolor}
\usepackage{balance}
\usepackage{siunitx}	% correct handling of SI units
\usepackage{comment}
\usepackage{cite}
\usepackage{makecell}

\ifCLASSOPTIONcompsoc
 \usepackage[caption=false,font=normalsize,labelfont=sf,textfont=sf]{subfig}
\else
 \usepackage[caption=false,font=footnotesize]{subfig}
\fi

\usepackage[normalem]{ulem}

\begin{document}

\title{Experimental Characterization of Air-to-ground Propagation at mm-Wave Frequencies in Dense Urban Environment}

\author{\IEEEauthorblockN{
Enrico M. Vitucci\IEEEauthorrefmark{1},   % 1st author, 1st affiliations
Vasilii Semkin\IEEEauthorrefmark{2}\IEEEauthorrefmark{3},   % 2nd author, 2nd affiliations
Maximilian J. Arpaio\IEEEauthorrefmark{1},    % 3rd author, 3rd affiliations
Marina Barbiroli\IEEEauthorrefmark{1},      % 4th author, 4th affiliations
\\
Franco Fuschini\IEEEauthorrefmark{1},      % 4th author, 4th affiliations
Claude Oestges\IEEEauthorrefmark{3},      % 4th author, 4th affiliations
and Vittorio Degli-Esposti\IEEEauthorrefmark{1}      % 4th author, 4th affiliations
}                                     % ...
%\\

\IEEEauthorblockA{\IEEEauthorrefmark{1} % 1st affiliations
Department of Electrical, Electronic, and Information Engineering “Guglielmo Marconi” (DEI), \\CNIT, University of Bologna, 40126 Bologna, Italy }
\IEEEauthorblockA{\IEEEauthorrefmark{2}% 2nd affiliations
VTT Technical Research Centre of Finland Ltd., 02150 Espoo, Finland, vasilii.semkin@vtt.fi}
\IEEEauthorblockA{\IEEEauthorrefmark{3}% 4th affiliations
 ICTEAM, Universit\'e catholique de Louvain, 1348 Louvain-la-Neuve, Belgium, claude.oestges@uclouvain.be}  

}



\maketitle

\begin{abstract}
In the present study, a measurement setup utilizing mm-wave transceivers with steerable directive antennas, mounted on both a customized UAV and a ground station has been used to study Air-to-Ground (A2G) radio links and, more generally, full-3D mm-wave propagation in urban environment. We evaluate the double-directional characteristics of the channel by rotating the antennas, deriving Power-Angle Profiles at both link ends. Preliminary results provide useful understanding of A2G propagation, e.g. the influence of the antenna tilt angles, or the mechanisms allowing for the signal to propagate from street canyons to the air. 
\end{abstract}

\vskip0.5\baselineskip

\section{Introduction}

The use of low-altitude, Unmanned Aerial Vehicles (UAV) in wireless communication, safety and sensing applications has been proposed by several researchers in recent years, and identified as one of the solutions for 5G key scenarios~\cite{zeng_commag, he_drone_PPDR}. Important advantages of UAV-aided solutions are their flexibility and the absence of a fixed infrastructure, a particularly attractive characteristic for temporary, on-demand services and for disaster-recovery applications. One of the prerequisites for the realization of UAV-aided wireless systems is the availability of reliable channel models for a large variety of environments, frequencies and UAV heights.

Several experimental investigations have been carried out for the characterization of air-to-ground (A2G) propagation, especially in the last few years~\cite{khawaja_survey}. Among the recent studies, several of them addressed rural or open-field propagation~\cite{cui_multi_freq_A2G_meas, khawaja_uwb_a2g} and propagation in university campuses and sub-urban areas ~\cite{cai_a2g_lte, qiu_uav_meas, cui_uwb_a2g, gauger_drone_mimo}. Only a few studies have addressed A2G propagation in actual urban areas, probably due to the inherent difficulties in getting authorizations to fly on densely populated zones and/or close them to public during measurements~\cite{zeleny_meas_low_elev, bucur_largescale_prop_uav, lopez_fading_uav}. In such studies, the analysis is mainly focused on large-scale parameters such as path-loss, fading statistics and spatial correlations. Existing studies – including those carried out in university campuses - are limited to UHF or sub-mmWave frequencies, and to our best knowledge, none of them investigated the double-directional characteristics of the A2G channel, probably due to the problems related to mounting and operating directive antennas or large MIMO transceivers on the UAV.

The present study is aimed at filling up some knowledge gaps in the field, and in particular the characterization of low-altitude A2G propagation in urban areas, where the presence of buildings has a strong impact on propagation, with focus on the channel’s directional characteristics at two mm-wave frequencies, 27 and 38 GHz, that are quite popular for having been recently allocated to 5G systems, although only 27 GHz results will be reported in the results section for the sake of brevity. Directive antennas at mm-wave frequencies can be relatively small and can help overcome power-budget limitations typical of UAV-aided communications. However, a thorough knowledge of the directional characteristics of the channel at both link-ends and advanced beam-steering techniques will be required to design and operate such systems.

The measurement setup of the present work is composed of a custom quadcopter equipped with a GPS plus Real-Time Kinematic (RTK) localization/navigation system, a directive horn antenna and a mm-wave portable spectrum-analyzer. An Ultra-Wide Band transceiver is also mounted on the drone to measure the channel’s time-domain characteristics, although in the lower band of 3.1-5.3 GHz. The ground station consists of the specular link-end with a directive antenna, a mm-wave generator and an UWB transceiver. Directive antennas are rotated in the azimuth or elevation planes at the ground station using a rotating positioner, while an automatic gimbal is used on the drone for 3D steering capabilities.

After the description of the measurement setup, some preliminary results are reported in the paper for a few cases of interest, including the Power-Angle Profiles (PAP) at both the ground station and at the drone in a street canyon for different drone heights. 

Besides being important for the design of future UAV-aided wireless systems, this setup and these measurements can also be of great interest for the study of full 3D mm-wave urban propagation, as the drone can mimic a base station or a user equipment that might be located anywhere from ground level to the top of the highest buildings or at mid-air on a cable within the urban environment, to allow for the maximum positioning flexibility.

\section{UAV-to-Ground Measurement Setup} \label{sec:setup}

Air-to-ground channel measurements have been carried out in urban environment at both millimeter and UWB frequencies, although only mm-wave results at 27 GHz will be reported in Section III for brevity.  

\begin{figure}[t]
\centering
\includegraphics[width=1.02\columnwidth]{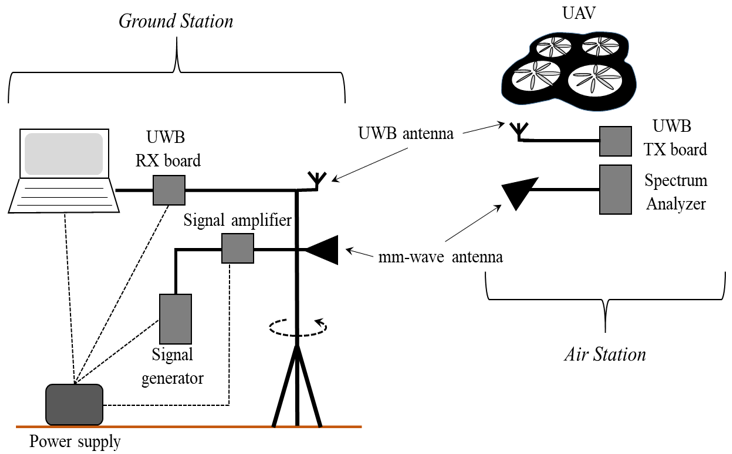}
\caption{Measurement set-up.}
\label{fig:setup}
\end{figure}

The experimental equipment was made of an air- and a ground-station (Fig.~\ref{fig:setup}), shortly outlined in the following sub-sections. Some information about the UAV piloting features and the software tools for data processing are also included.

\begin{figure}[t]
\centering
\includegraphics[width=1.02\columnwidth]{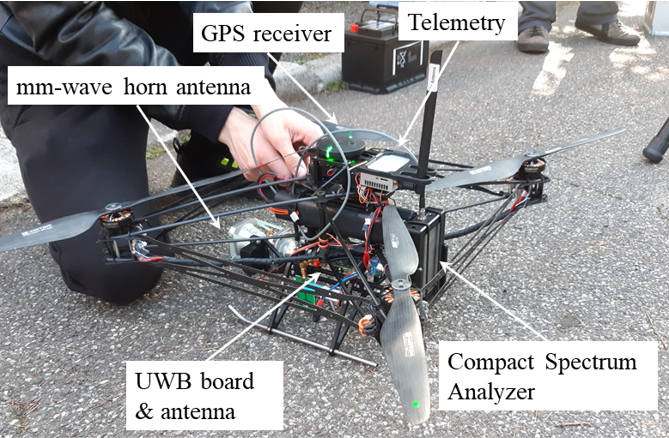}
\caption{Air station.}
\label{fig:air_station}
\end{figure}

\subsection{Air-station}
\label{subsec:air}
The UAV clearly represents the backbone of the air-station. It is a quadcopter drone with a maximum weight of 4 kg (including measurement setup), specifically conceived and customized for wireless channel measurement (Fig.~\ref{fig:air_station}). The four rotating blades are supplied by 380~W electric motors placed at the corners of the UAV structure. Both the blades and the drone frame are made of carbon fiber to reduce its weight. A metal slid is present on the bottom side to let the UAV safely rest on ground, whereas the top side hosts the telemetry and the remote-control units and the GPS receiver. Real Time Kinematic (RTK) can be enabled in order to increase positioning accuracy. Electrical motors and the other devices are supplied by 6s Li-Ion batteries, corresponding to a maximum flight time equal to about 20 minutes.

On-board equipment for millimeter-wave communications  consists of a SAF Tehnika J0SSAP14 compact spectrum analyzer (SA) operating in the 24-40~GHz band with a sweep speed of 0.5~s at a 100~MHz span~\cite{spectrumcompact}. Received signals are provided to the SA by either a conical horn or an omnidirectional antenna, with gain respectively equal to 21dB and 3dB. The elevation of the directive antenna can be automatically set and controlled by means of a servo-control.
UWB wireless signals in the 3-5~GHz band are also radiated at the UAV through an omnidirectional antenna fed by a Time Domain PulseON P410 transmitting board.
The total weight of the mm-wave equipment amounts to about 0.5~kg, whereas it is of a few tens of grams for the UWB setup. Both the mm-wave SA and the UWB transmitter are powered by dedicated batteries to save the drone battery life.

\subsection{Ground-station}
\label{subsec:ground}

At ground level, the millimeter-wave link is completed by a SAF Tehnika J0SSAG14 compact signal generator (SG), with a maximum output power equal to 5dBm~\cite{spectrumcompact}. The intensity of the transmitted signal is then boosted by means of a SAGE Ka-band power amplifier having a gain equal to about 20dB.  Depending on the specific need, the amplifier can feed the same antennas already introduced for the air station. In order to carry out directional channel measurements, the horn antenna can be steered in the azimuth domain by means of a software controlled Yaesu g-450 rotating positioner.

An omnidirectional UWB antenna connected to a second PC-controlled PulseON P410 board is also part of the ground segment, to get and store the signal coming from the UWB transmitter placed on the UAV.

A power inverter connected to a 24V car battery provides the ground equipment with the necessary power supply.

\subsection{Control Firmware and Data Processing}
\label{subsec:data_proc}

Although a pilot must be always present in clear sight of the UAV for safety reasons, the UAV flights have been planned beforehand by means of the QGroundControl application~\cite{groundcontrol}. Every flying mission basically consists of a sequence of “waypoints” the UAV moves on and a list of target points to steer the horn antenna towards (when required, of course). At the end of each mission, the flight data – including the temporal variations in the UAVs position and steering direction – are available in a specific telemetry file.

In order to match the different UAV position/pointing with the measured received signal strength intensity, both the SA and the drone control boards are synchronized to the Coordinated Universal Time.

Finally, propagation markers like power angle/delay profile, angle and delay spread, blockage loss, path loss exponent etc. can be extracted from the measured and telemetry data by means of ad-hoc Matlab scripts.

\section{Results and Discussion} \label{sec:results}

In this Section, we present preliminary results of the measurement campaign. For the sake of brevity, in this work we present only measurements obtained using the mm-Wave equipment, at a frequency of 27 GHz. In particular, in the following directional measurements at both air station and ground station are shown: Power-Angle Profiles (PAP) have been obtained by rotating the directional horn antennas in both azimuth and elevation angles, and such results have been analyzed in order to get useful insights on the dominant propagation mechanisms.
Two measurement scenarios are described below, located in an urban street canyon located in the town of Imola, Italy: the corresponding measurement setup (UAV height, antennas, etc.) is summarized in Table~\ref{tab:summary}. In both cases, a great deal of work has been done to get the necessary authorizations to fly with the custom drone on the urban area from both national and local Authorities, and to get permissions to close the streets, alert the residents and monitor the area during measurements.

\begin{table*}[t]
    \centering
    \caption{Summary of the measurement scenarios}
    \begin{tabular}[font=\medium]{|c|c|c|c|c|c|}
        \hline
        \makecell{Meas. Scenario} & \makecell{Freq. [GHz]} & \makecell{Tx Antenna} & \makecell{Tx Height, [m]} & \makecell{UAV Antenna} & \makecell{UAV Height, [m]} \\
        \hline
        \multirow{2}{*}{\makecell{N\textsuperscript{\underline{o}}1}} & \multirow{2}{*}{27} & \multirow{2}{*}{\makecell{Fixed (Omni)}} & \multirow{2}{*}{2} & \multirow{2}{*}{\makecell{Rotating Horn\\(Azimuth + Elevation)}} & 19 \\
        \cline{6-6}
        &  &  &  &  & 50\\
        \hline
        \multirow{2}{*}{\makecell{N\textsuperscript{\underline{o}}2}} & \multirow{2}{*}{27} & \multirow{2}{*}{\makecell{Rotating Horn\\(Elevation)}} & \multirow{2}{*}{2} & \multirow{2}{*}{\makecell{Fixed (Horn)}} & 19 \\
        \cline{6-6}
        &  &  &  &  & 50\\
        \hline
    \end{tabular}
    \label{tab:summary}
    
\end{table*}{}

\subsection{Measurement Scenario 1}

In this scenario, an omnidirectional TX antenna is used at the ground station, while a conical horn antenna is used at the RX air station. The ground station is placed on one side of a North-oriented urban street canyon, while the air station (drone) is hovering above the roof of a building on the opposite side of the street.
In order to get Power-Azimuth Profiles at the RX, the drone rotates from 0$^{\circ}$ to 360$^{\circ}$ keeping its position fixed, with angular steps of 20$^{\circ}$. The azimuth angle is assumed 0° when the antenna is pointing at North and increases clockwise. For each orientation, the drone keeps its angular position for 10 seconds, so that the measured data can be properly averaged in time. The transition from one angular position to the next one takes about 1 second, and the measured data acquired during the transition are discarded.

This measurement is then repeated for different tilt angles of the RX horn antenna, which is pointed downward using the servo control. 

\begin{figure}[!t]
\centering
\includegraphics[width=0.85\columnwidth]{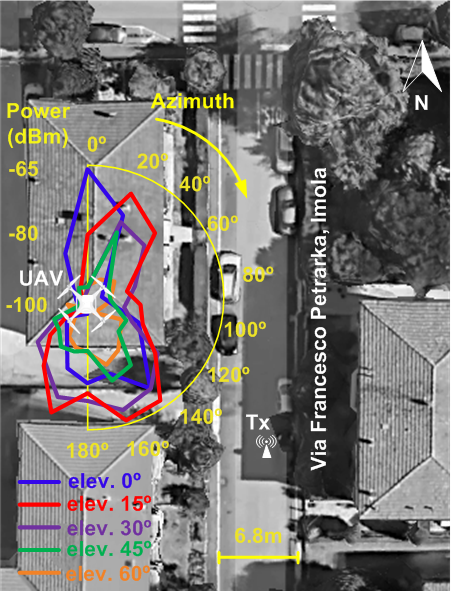}
\caption{Top-view of the street-canyon scenario and Power-Azimuth Profile at RX with drone on building at very low altitude (19 m, about 5 m above rooftop), for different tilt angles.}
\label{fig:power_az1}
\end{figure}

\begin{figure}[!t]
\centering
\includegraphics[width=0.85\columnwidth]{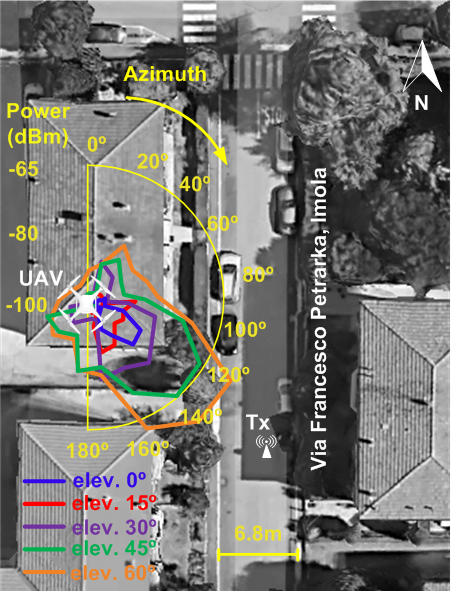}
\caption{Top-view of the street-canyon scenario and Power-Azimuth Profile at RX with drone on building at medium altitude (50 m), for different tilt angles.}
\label{fig:power_az2}
\end{figure}

Results are shown in Fig.~\ref{fig:power_az1} and~\ref{fig:power_az2} for two different heights of the drone: a very low altitude, few meters above the average building height (19 m), and a medium altitude (50 m). 
In the plots, we see a representation of the PAPs on the 2D map of the street canyon: each plot, with a different color, corresponds to a different tilt angle of the RX antenna, from 0$^{\circ}$ to 60$^{\circ}$, with a step of 15$^{\circ}$.

Looking at Fig.~\ref{fig:power_az1}, we see that optimum tilt angle is 15° (red curve), while performance is highly degraded when the drone antenna is pointing downward at 60° and 45° (yellow and green curve), because in such cases the RX antenna has maximum gain when pointing towards the roof. In the yellow curve (60$^{\circ}$ tilt), power is on average 25 dB lower than in red curve (15$^{\circ}$ tilt). 

In the same case (Fig.~\ref{fig:power_az1}), for the lower tilt angles we also observe strong “sidelobes” at 20$^{\circ}$ and 200$^{\circ}$ in the PAPs, corresponding to arrival directions other than the LOS path: these are probably caused by specular reflection on the façades of surrounding buildings (some of them are higher than the one where the drone is flying on).
In Fig.~\ref{fig:power_az2} (drone flying at 50 meters) we see the opposite behaviour: the optimum tilt angle is 60$^{\circ}$, while the performance is poor for low elevation angles. In such a case, propagation is dominated by the LOS path, and there are no evident “sidelobes” in the PAP.

\subsection{Measurement Scenario 2}

\begin{figure}[t]
\centering
\includegraphics[width=1.02\columnwidth]{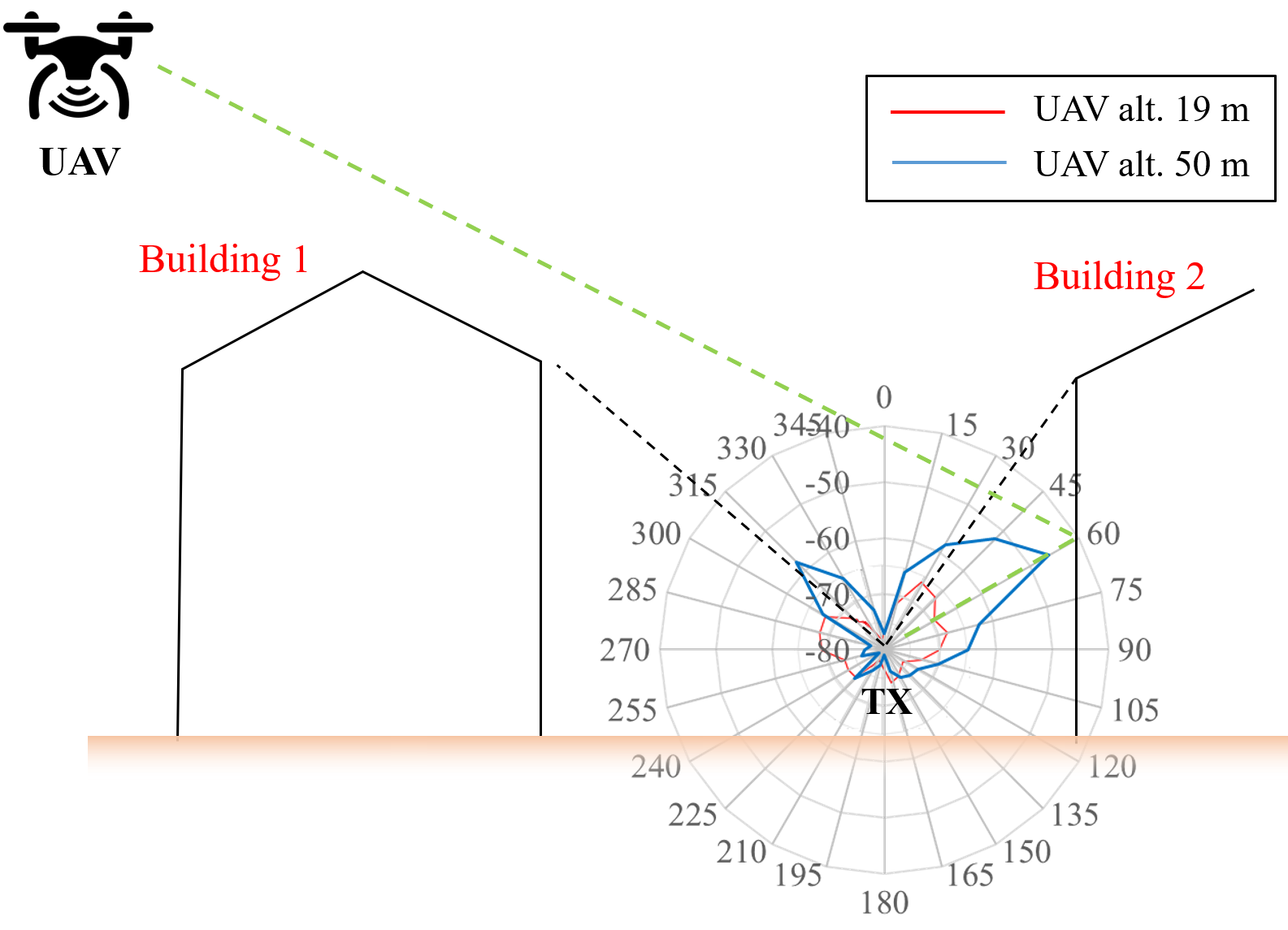}
\caption{Power-Elevation Profiles at TX in the “Air-to-street” NLOS scenario, for 2 different drone altitudes: 19 m (red curve) and 50 m (blue curve).}
\label{fig:power_elevation}
\end{figure}

In this scenario, represented in Fig.~\ref{fig:power_elevation}, the drone is hovering beyond the street canyon at a fixed position, with the RX horn antenna oriented towards the center of the street canyon, while the ground station with TX is placed on the farther side of the street canyon: the LOS path is therefore obstructed by the left side of the street canyon (building 1, see Fig.~\ref{fig:power_elevation}).

In this case, the horn conical antenna is also used at the TX ground station: in particular, the antenna is fixed to a horizontal mast pole which rotates during the measurement, in order to get a Power-Angle Profile at the TX on the Vertical plane, i.e. a Power-Elevation Profile (PEP).

Results are shown in Fig.~\ref{fig:power_elevation}, where the red curve corresponds to the low-altitude case (19 m), and blue curve to medium-altitude case (50 m). In the plots, the 0$^{\circ}$ angle corresponds to the TX antenna pointed vertically towards the sky, and the angle increases clockwise, i.e. when the antenna rotates towards the building behind the ground station (Building 2), with 15$^{\circ}$ step.

Looking at the blue curve (50 m altitude), we can clearly identify a dominant contribution at 60° departure angle: this contribution appears to be compatible with a specular reflection (green dashed line in Fig.~\ref{fig:power_elevation}) on the main façade of Building 2, i.e. the building behind the ground station. This contribution is about 10 dB stronger than the one provided by the minimum-length path, corresponding to a diffraction on the horizontal roof edge (315$^{\circ}$ departure angle in blue curve). Reflection/scattering on the building behind TX appears then to be one of the main mechanisms for the signal to propagate from the street-canyon to the air (and viceversa), in agreement with previous studies~\cite{degli-esposti_roof2street}.

Looking at the red curve (19 m altitude), we notice that the power-levels are on average 15 dB lower than in the previous case: this is due to less favourable conditions for the signal to propagate from the street canyon to the air. In such a case, the geometry is not compatible with specular reflections on the building walls surrounding TX: the PEP plot has two relative maxima with similar intensity, at 35$^{\circ}$ and 310$^{\circ}$ (black dashed lines in Fig.~\ref{fig:power_elevation}), which appears to be caused by diffraction on roof edges, or scattering from roof drain pipes.

\section{Conclusions}
\label{sec:conclusion}

In the present study, a measurement setup utilizing mm-wave transceivers with steerable directive antennas, mounted on both a customized UAV and a ground station has been used to study Air-to-Ground radio links in urban environment. Results provide useful insights on full-3D, mm-wave urban propagation, like the influence of the antenna azimuth and elevation angles at the UAV, and the main propagation mechanisms allowing the radio signal to propagate from urban street canyons to the air. In future work, further measurements will be presented and analyzed for both 27 and 38 GHz and for different positions of drone and ground station within urban environment. Moreover, the results provided by the UWB setup will be analyzed, including time-domain results, and compared to mm-Wave frequencies.

\section*{Acknowledgment}
Authors would like to thank AslaTech for taking care of UAV customization, permission to fly and flight safety throughout the measurement activity.
This work is supported in part by the Italian Ministry of University and Research (MUR) under the programme “Dipartimenti di Eccellenza (2018–2022)—Precision Cyberphysical Systems (P-CPS)”, and by the EU COST action CA15104 “Inclusive Radio Communications (IRACON)”.\\The work of V. Semkin is supported in part by the Jorma Ollila grant and by the Academy of Finland. The work of C. Oestges is supported by MUSEWINET project under the Belgian Science Foundation FRS-FNRS (Fonds de la Recherche Scientifique) EOS programme.

\bibliographystyle{IEEEtran}
\bibliography{bibliography}

\end{document}